\documentclass[conference]{IEEEtran}
\usepackage{epsfig}
\usepackage{times}
\usepackage{float}
\usepackage{afterpage}
\usepackage{amsmath}
\usepackage{amssymb}
\usepackage{amstext}
\usepackage{bm}
\usepackage{latexsym}
\usepackage{color}
\usepackage{array,algpseudocode}
\usepackage{makecell}
\usepackage{diagbox}
\usepackage{amsthm}
\usepackage{amssymb} 
\usepackage{graphicx}
\usepackage{caption}
\usepackage{pstricks}
\usepackage{booktabs}
\usepackage{enumerate}
\usepackage[ruled]{algorithm}
\usepackage{cite}
\usepackage{subcaption}
\usepackage{flushend}
\usepackage[left = 0.72in, right=0.72in, top=0.75in, bottom=1.05in]{geometry}
\usepackage{bbm}

\usepackage[normalem]{ulem}
\usepackage{url}

\newcommand{\cO}{\mathcal{O}}
\newcommand{\comment}[1]{}

\usepackage{mathbbol}

\def\mindex#1{\index{#1}}

\def\sq{\hbox{\rlap{$\sqcap$}$\sqcup$}}
\def\qed{\ifmmode\sq\else{\unskip\nobreak\hfil
\penalty50\hskip1em\null\nobreak\hfil\sq
\parfillskip=0pt\finalhyphendemerits=0\endgraf}\fi\medskip}

\long\def\defbox#1{\framebox[.9\hsize][c]{\parbox{.85\hsize}{%
\parindent=0pt
\baselineskip=12pt plus .1pt      % STYLE
\parskip=6pt plus 1.5pt minus 1pt % CHANGES
 #1}}}

\long\def\beginbox#1\endbox{\subsection*{}%
\hbox{\hspace{.05\hsize}\defbox{\medskip#1\bigskip}}%
\subsection*{}}

\def\endbox{}

\newsavebox{\junk}
\savebox{\junk}[1.6mm]{\hbox{$|\!|\!|$}}

\def\bfmath#1{{\mathchoice{\mbox{\boldmath$#1$}}%
{\mbox{\boldmath$#1$}}%
{\mbox{\boldmath$\scriptstyle#1$}}%
{\mbox{\boldmath$\scriptscriptstyle#1$}}}}

\def\bfmY{\bfmath{Y}}

\def\bfmhhaY{\bfmath{\hhaY}} %\widehat{\widehat{Y}}}}
\def\bfmhhaY{\hbox to 0pt{$\widehat{\bfmY}$\hss}\widehat{\phantom{\raise 1.25pt\hbox{$\bfmY$}}}}

\def\til={{\widetilde =}}

 \def\FRAC#1#2#3{\genfrac{}{}{}{#1}{#2}{#3}}

\def\ddtp{{\mathchoice{\FRAC{1}{d^{\hbox to 2pt{\rm\tiny +\hss}}}{dt}}%
{\FRAC{1}{d^{\hbox to 2pt{\rm\tiny +\hss}}}{dt}}%
{\FRAC{3}{d^{\hbox to 2pt{\rm\tiny +\hss}}}{dt}}%
{\FRAC{3}{d^{\hbox to 2pt{\rm\tiny +\hss}}}{dt}}}}

\def\average#1,#2,{{1\over #2} \sum_{#1}^{#2}}

\def\eye(#1){{\bf(#1)}\quad}

\def\eq#1/{(\ref{e:#1})}

\newcommand{\beqn}[1]{\notes{#1}%
\begin{eqnarray} \elabel{#1}}

\newcommand{\eeqn}{\end{eqnarray} }

\newcommand{\beq}[1]{\notes{#1}%
\begin{equation}\elabel{#1}}

\newcommand{\eeq}{\end{equation}}

\def\bdes{\begin{description}}
\def\edes{\end{description}}

\newcounter{rmnum}

\newcounter{anum}

{\end{list}}

\def\ass(#1:#2){(#1\ref{#1:#2})}

\def\ritem#1{
\item[{\sf \ass(\current_model:#1)}]
}

\newenvironment{recall-ass}[1]{%
\begin{description}
\def\current_model{#1}}{
\end{description}
}

\long\def\comment#1{}

\newfont{\bb}{msbm10 scaled 1100}

\let\svthefootnote\thefootnote
\newcommand\blankfootnote[1]{%
  \let\thefootnote\relax\footnotetext{#1}%
  \let\thefootnote\svthefootnote%
}

\allowdisplaybreaks
\begin{document}
\title{Blind Source Separation-Enabled Joint Communication and Sensing in IBFD MIMO Systems
}
\iffalse
\author{\IEEEauthorblockN{ Siyao Li}
\IEEEauthorblockA{\textit{Department of Electrical Engineering and Computer Science} \\
\textit{Embry-Riddle Aeronautical University}\\
Daytona Beach, FL, USA \\
lis14@erau.edu}
\and
\IEEEauthorblockN{ Conrad Prisby}
\IEEEauthorblockA{\textit{Department of Electrical Engineering and Computer Science} \\
\textit{Embry-Riddle Aeronautical University}\\
Daytona Beach, FL, USA \\
prisbyc@my.erau.edu}
\and
\IEEEauthorblockN{ Thomas Yang}
\IEEEauthorblockA{\textit{Department of Electrical Engineering and Computer Science} \\
\textit{Embry-Riddle Aeronautical University}\\
Daytona Beach, FL, USA \\
yang482@erau.edu}
}
\fi
\author{
	\IEEEauthorblockN{Siyao Li,  %\IEEEauthorrefmark{1}, 
    Conrad Prisby, % \IEEEauthorrefmark{1},
     %William Reimer, %\IEEEauthorrefmark{1},
            Thomas Yang %\IEEEauthorrefmark{1}
	}
\IEEEauthorblockA{\IEEEauthorblockA{ Department of Electrical Engineering and Computer Science, 
\\ Embry-Riddle Aeronautical University, Daytona Beach, FL, USA}
\IEEEauthorblockA{E-mail: lis14@erau.edu, prisbyc@my.erau.edu,  yang482@erau.edu}
} }

\maketitle

\begin{abstract}

This paper addresses the challenge of joint communication and sensing (JCAS) in next-generation wireless networks, with an emphasis on in-band full-duplex (IBFD) multiple-input multiple-output (MIMO) systems. Traditionally, self-interference (SI) in IBFD systems is a major obstacle to recovering the signal of interest (SOI). Under the JCAS paradigm, however, this high-power SI signal presents an opportunity for efficient sensing. Since each transceiver node has access to the original SI signal, its environmental reflections can be exploited to estimate channel conditions and detect changes, without requiring dedicated radar waveforms. We propose a blind source separation (BSS)-based framework to simultaneously perform self-interference cancellation (SIC) and extract sensing information in IBFD MIMO settings. The approach applies the Fast Independent Component Analysis (FastICA) algorithm to separate the SI and SOI signals while enabling simultaneous signal recovery and channel estimation. Simulation results confirm the framework’s effectiveness, showing improved sensing and communication performance as signal frame size increases.

%BSS enables efficient and accurate channel estimation, significantly enhancing the performance of JCAS systems. 

%This work addresses the joint communication and sensing (JCAS) challenge using orthogonal frequency division multiplexing (OFDM) signals without prior knowledge of the source signals. In this context, passive radar receivers capture reflected or scattered communication signals from multiple objects to estimate channel conditions or detect environmental changes, without relying on dedicated radar signals. However, the absence of direct access to the original transmitted signal introduces significant challenges for achieving real-time and accurate sensing. To overcome this, we leverage Blind Source Separation (BSS) techniques that exploit the structure and statistical properties of the received signals, enabling efficient and accurate sensing without requiring prior knowledge of source signals.

\end{abstract}


\section{Introduction}
\label{sec:intro}

The rapid growth of mobile data traffic and the shift toward millimeter-wave (mmWave) communication necessitate an evolution in wireless technologies to meet increasing bandwidth demands \cite{heath2016overview}. However, spectrum scarcity remains a major challenge, limiting the capacity to support high data rates, as reflected in the soaring cost of spectrum allocation \cite{Belgium}. 
To enhance spectral efficiency and resource utilization, in-band full-duplex (IBFD) multiple-input multiple-output (MIMO) wireless communications have gained significant attention \cite{kolodziej2021band, alves2020full},  which allows a wireless transceiver to transmit and receive data simultaneously using the same frequency band. 
Nevertheless, a major challenge for IBFD systems is managing self-interference (SI), the leakage of a node’s own transmitted signal into its receiver chain, which can overshadow the weaker signal of interest (SOI).

The integration of joint communication and sensing (JCAS) is particularly aligned with emerging military requirements for FutureG and beyond \cite{wen2024survey}. Modern defense operations demand resilient, spectrum-efficient systems capable of simultaneously maintaining covert, jam-resistant communications while providing situational awareness. By reusing the SI signal as a sensing probe, IBFD JCAS systems reduce electromagnetic signature, conserve bandwidth, and enable real-time detection of environmental changes such as adversarial jamming, drone surveillance, or battlefield mobility \cite{smida2024band}. These dual-use capabilities highlight the timeliness of JCAS for defense-oriented applications, where spectrum agility, stealth, and multifunctionality are critical (e.g., see \cite{EnableJCAS,li2022capacity,li2023capacity,Fang2023JCAS,li2024capacity,Song2024,ISAC-6G,ISAC-IT} and references therein).
%While conventional designs consider SI an impediment, the emerging paradigm of joint communication and sensing (JCAS) proposes a paradigm shift. Rather than treating SI as detrimental, JCAS reinterprets it as a valuable source for environmental sensing \cite{smida2024band}. Since the SI signal is fully known to the transmitter node, it can be utilized to extract environmental reflections and estimate channel conditions without requiring separate sensing waveforms. This dual-functionality opens up promising opportunities for low-cost, high-efficiency spectrum usage in next-generation wireless systems (e.g., see \cite{EnableJCAS,Fang2023JCAS,BBP:TIT2023,ISAC-6G,ISAC-IT} and references therein).

Existing approaches to sensing often rely on deterministic reference signals  \cite{wei20225g,bekkerman2006target}, while communication employs random modulation schemes such as QAM or BPSK. From an information-theoretic perspective, treating these functionalities separately is suboptimal \cite{lapidoth1998reliable}. Instead, integrating sensing with communication at the physical layer has the potential to significantly improve efficiency and reduce hardware and energy overhead.

Recent research has increasingly focused on leveraging blind source separation (BSS) techniques for JCAS within IBFD MIMO frameworks. A comprehensive survey  outlines BSS applications in adaptive wireless systems, demonstrating its potential for handling overlapping signals in MIMO-based cognitive radios and spectrum sensing contexts \cite{Luo2018}.  A sparsity-enhanced source separation scheme is proposed in \cite{Jin2022} to mitigate aliasing in joint communication and radar scenarios. Fouda et al. \cite{Fouda2021} further explored the design of BSS architectures tailored for full-duplex systems, emphasizing the challenges associated with maintaining computational efficiency while managing strong SI signals. Barneto \cite{Baquero2022} provides insights into waveform and hardware integration strategies, offering a foundational understanding of signal co-design in cellular JCAS systems. %\textcolor{blue}{Beyond civilian networks, these advancements could also benefit the military by allowing UAVs to communicate reliably while using radar-like sensing to detect threats in contested areas.}
 %Smida et al. \cite{smida2024band} provide a broad perspective on the use of IBFD MIMO systems for simultaneous communication and sensing, identifying BSS as a key enabling technique for dynamic, dual-purpose wireless environments.

\paragraph*{Contributions} Despite these advances, few works have examined the use of BSS in settings where the SI signal is known and can be repurposed for environmental sensing. This paper addresses this gap by applying classical BSS methods in a scenario optimized for practical JCAS deployment. Specifically, we apply the Fast Independent Component Analysis (FastICA) algorithm  \cite{bingham2000fast}  to decompose the received signal into its constituent sources, enabling both self-interference cancellation (SIC) and environmental sensing.  Our contributions thus extend existing work by demonstrating a computationally efficient approach to JCAS that balances spectral efficiency, channel estimation accuracy, and system simplicity. 
We validate the proposed framework through simulations under practical conditions, evaluating performance based on estimation accuracy, signal separation, and convergence behavior. Results demonstrate that our BSS-based solution provides effective joint sensing and communication with minimal overhead, making it a strong candidate for future full-duplex wireless systems.

\paragraph*{Organization} The remainder of this paper is organized as follows. Section \ref{sec:system} introduces the system model. Section \ref{sec:BSS} presents the BSS algorithm. Section \ref{sec:simulation} evaluates the performance with numerical examples. Section \ref{sec:conclusion}
 concludes this work and discusses future endeavors. 
\paragraph*{Notations} 
  We use $\mathbb{E}[X]$ to represent the expectation of the random variable $X$. Let bold letter $\bf{X}$ denote a vector or matrix.  $\| \cdot \|^2$ represents the Frobenius norm. $\mathcal{N}(\mu, \sigma^2)$ denotes Gaussian distribution with mean $\mu$ and variance $\sigma^2$. ${\bf M}^{n\times m}$ represents a generic matrix with $n$ rows and $m$ columns.

\begin{figure}[t]
\centerline{ \includegraphics[width=1.05\linewidth]{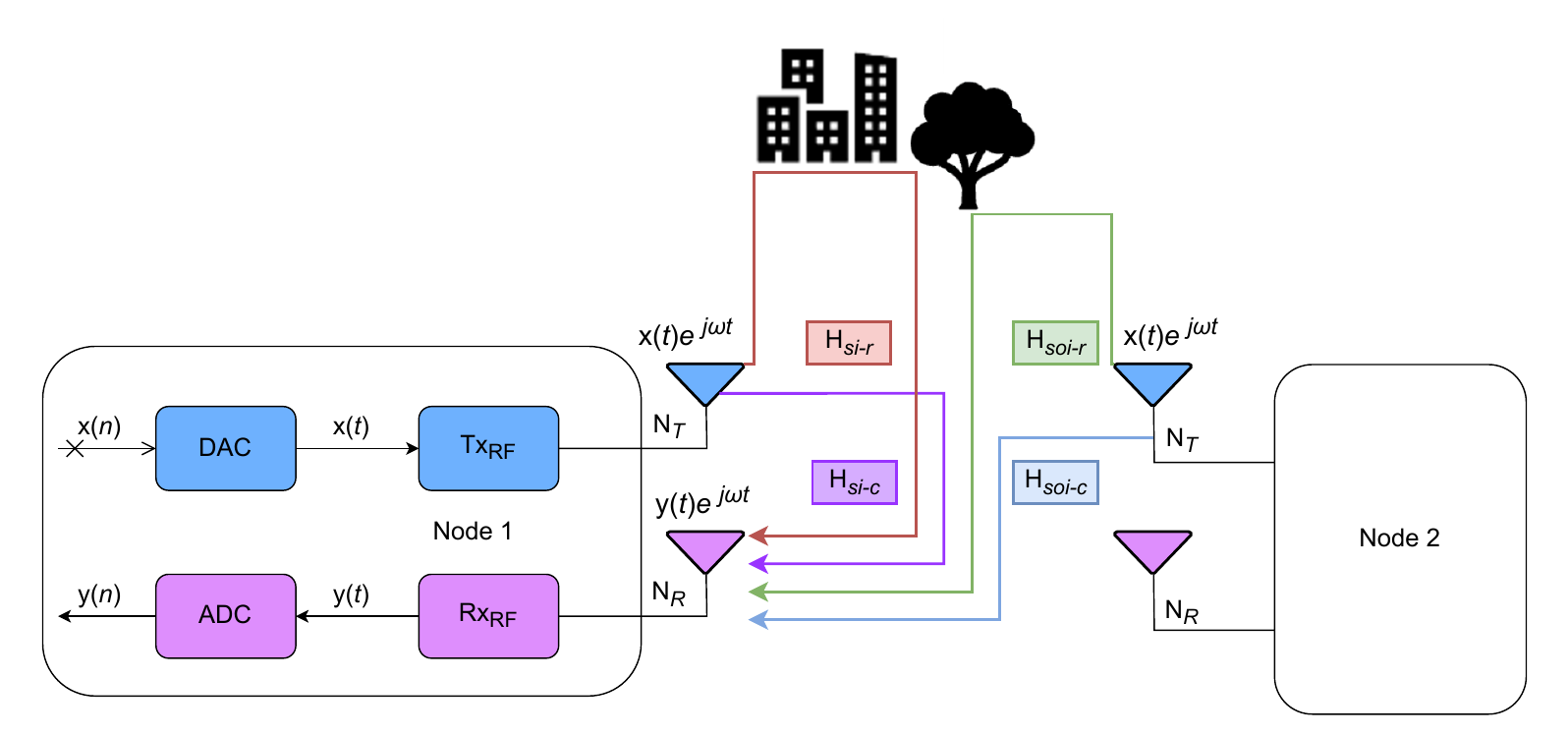} \hspace{.0cm}  }
\caption{IBFD system model.}
     \label{fig:model}
	\vspace{-.55cm}
\end{figure}

\section{System Model}
\label{sec:system}
As illustrated in Fig. \ref{fig:model}, we consider an IBFD MIMO system comprising two transceiver nodes, each equipped with  $n$  transmit antennas and  $m$  receive antennas. Both nodes simultaneously transmit and receive signals over the same frequency band, enabling full-duplex communication. 
The received signal at each node is composed of multiple components: (1) the SOI, which includes both direct-path and reflected components from the remote node; and (2) the SI signal, which includes direct leakage between the local transmit and receive antennas and reflections from the environment. %At each node, the received signal is a superposition of multiple components, including the SOI, which propagates through the direct path between the two nodes and experiences reflections from surrounding objects, as well as the SI signal, which travels both directly between the transmit and receive antennas of the same node and through multipath reflections within the environment. 
%multiple versions of the signals of interest from the other node and multiple versions of the self-interference signals from its own transmit antennas.
%Thus, the received signal is a mixture of the signals of interest, the self-interference signals, and the sensing echo.
Let $ R_l(k)$ denote the frequency-domain received signal at the $l$-th receive antenna at time $k$. It is expressed as: %The received signal at the $l$-th receiver can be written in the frequency domain as follows.
\begin{align}
    R_l(k) &= \sum_{i=1}^{n} ( H_\text{si-c}^i(k)  + H_\text{si-r}^i(k) ) S_\text{si}^i(k) %+ \sum_{i=1}^{n} H_\text{soi}^i(k) S_\text{soi}^i(k)
    \notag
    \\ &+ \sum_{i=1}^{n} ( H_\text{soi-c}^i(k) + H_\text{soi-r}^i(k) )S_\text{soi}^i(k) + N(k), \ l \leq m.
    \label{eq:Rlk}
\end{align}
 Here, $N(k) \sim \mathcal{N}(0, \sigma^2)$ is the additive Gaussian noise; $H_\text{si-c}^i(k)$ and $H_\text{soi-c}^i(k)$ are the direct SI and SOI channels respectively; $H_\text{si-r}^i(k)$ and $H_\text{soi-r}^i(k)$ denote the reflected channels at the $i$-th transmit antenna the discrete time instant $k$; 
 $S_\text{si}^i(k)$ and  $S_\text{soi}^i(k)$ are the SI and SOI signals from the $i$-th transmit antenna respectively. Each channel matrix is assumed to have independent, identically distributed (i.i.d.)  entries. For convenience, we can drop the time index and  rewrite \eqref{eq:Rlk} in matrix form as 
 \begin{align}
    {\bf R} = ({\bf H}_\text{si-c} + {\bf H}_\text{si-r}) {\bf S}_\text{si} + ({\bf H}_\text{soi-c}  + {\bf H}_\text{soi-r} ) {\bf S}_\text{soi} + {\bf N},
    \label{eq:R}
\end{align}
where ${\bf H}_\text{si-c}, {\bf H}_\text{si-r}, {\bf H}_\text{soi-c}, {\bf H}_\text{soi-r} \in {\bf M}^{m \times n}$, ${\bf S}_\text{si}, {\bf S}_\text{soi} \in {\bf M}^{n \times 1}$, and ${\bf R} \in {\bf M}^{m \times 1}$. In particular, ${\bf H}_\text{si-c}$ is the direct self-interference channel, which is in general known to the receiver{\footnote{In practical IBFD systems, the direct SI channel is dominated by the line-of-sight path between the transmitter and receiver antennas of the same node, and primarily depends on the fixed and deterministic hardware characteristics of the transceiver. It can be estimated offline during system calibration or periodically updated using pilot signals. Since the transceiver layout remains unchanged, ${\bf H}_\text{si-c}$ remains relatively stable over time, allowing the receiver to effectively use it for self-interference cancellation techniques.}}.
In this work, we use ${\bf H}_\text{si-r}$ to characterize the reflected channel containing the sensing information. 
\iffalse
Consider the time-invariant backscatter channel with $P$ static targets{\footnote{Here, we consider the channel from the BS transmitter to the BS radar receiver, reflected by the targets.}}, whose response is given by 
\begin{align}
h\left( {\tau } \right) = \sum\limits_{p = 1}^P {h_p}
%{\bf{a}}} \left( {{\phi _p}} \right){{\bf{a}}^{\rm{H}}}\left( {{\phi _p}} \right)
\delta \left( {\tau  - {\tau _p}} \right). \label{channel_impulse_response}
\end{align}
For each target $p$ in~\eqref{channel_impulse_response}, $h_p$ models the complex radar reflectivity following a zero mean circularly-symmetric complex Gaussian distribution with variance $C_{h_p} \overset{\Delta}{=}\mathbb{E}[| {{h_p}}|^2] = \frac{{{\lambda ^2}\sigma _{p, {\rm{rcs}}}}}{{{{\left( {4\pi } \right)}^3}d_p^4}}$~\cite{richards2014fundamentals}, where  
$\lambda$ is the signal wavelength, ${\sigma _{p,{\rm{rcs}}}}$ is the radar cross section (RCS) and $d_p$ is the relative distance between the $p$-th target and the node. Moreover, $\tau_p = \frac{2 d_p}{c}$ is the round-trip delay (time of flight), where $c$ denotes the speed of light.
The $k$-th diagonal element of ${\bf H}_\text{soi-r}$ is 
\begin{align}
    h_k = \sum\limits_{p = 1}^P {h_p} e^{j 2\pi \frac{k-1}{T}{\tau _p}}.
\end{align}
\fi
 Note that ${\bf S}_\text{si}$ is the transmitted signal from the same node, implying that ${\bf S}_\text{si}$ is known to the receiver. We can further rewrite \eqref{eq:R}   in the matrix form:
\begin{align}
\label{eq:matrix-system}
    \begin{bmatrix}
       {\bf S}_\text{si}
       \\
       {\bf R}
    \end{bmatrix}  =  {\bf H}   \begin{bmatrix}
       {\bf S}_\text{si}
       \\
       {\bf S}_\text{soi}
    \end{bmatrix} + \begin{bmatrix}
       {\bf 0}
       \\
       {\bf N}
    \end{bmatrix},
\end{align}
where $ {\bf H}$ is a composite mixing matrix defined as:
%\begin{subequations}
\begin{align}
   % {\bf S} = \begin{bmatrix}
   %    {\bf S}_\text{si},
   %    \\
   %    {\bf S}_\text{soi}
   % \end{bmatrix} \in \mathbb{C}^{2n\times 1}, \ 
     {\bf H} = \begin{bmatrix}
      {\bf I} & {\bf 0}
       \\
   {\bf H}_\text{si}  & {\bf H}_\text{soi} 
    \end{bmatrix} \in {\bf M}^{2m\times 2n},
    \label{eq:R-matrix}
    \\
    {\bf H}_\text{si} =  {\bf H}_\text{si-c} + {\bf H}_\text{si-r},
    \\
    {\bf H}_\text{soi}  = {\bf H}_\text{soi-c} + {\bf H}_\text{soi-r},
    \label{eq:H-SOI}
\end{align}
%\end{subequations}
and ${\bf I}$ is the identity matrix of size $m \times n$. 
This formulation in  \eqref{eq:matrix-system}  represents a canonical BSS model \cite{independenthandbook}, where known and unknown sources are mixed via partially known channels. %\textcolor{blue}{The computational complexity of FastICA is primarily $\mathcal{O}(nm^2 + m^3)$, where $n$ is the number of data samples and $m$ is the number of independent components, arising from the iterative optimization of the separation matrix, while matrix inversion typically has a complexity of $\mathcal{O}(n^3)$ using algorithms such as Gaussian elimination.} 
The key idea in this paper is to exploit knowledge of the SI signal ${\bf S}_\text{si}$ and its direct channel ${\bf H}_\text{si-c}$ to estimate both the reflection component ${\bf H}_\text{si-r}$ and the unknown communication signal ${\bf S}_\text{soi}$. 

%Equation  \eqref{eq:matrix-system} represents the well-known BSS signal model \cite{independenthandbook}.
%\subsection{OFDM Signaling for Channel Estimation}

%\subsection{Problem Formulation}

\section{BSS Based Channel Estimation}
\label{sec:BSS}
We now describe the application of the  FastICA algorithm \cite{bingham2000fast} to separate the SI and SOI components from the observed signals. For simplicity, we assume $m = n$, meaning each node has an equal number of transmit and receive antennas. %In this section, we apply the FastICA  algorithm \cite{bingham2000fast}, a classical BSS technique,  to the formulated IBFD MIMO system in \eqref{eq:matrix-system}. For analytical simplicity, we assume $m = n$, meaning that each node has an equal number of transmit and receive antennas.
%While communication performance in IBFD MIMO systems has been extensively studied (see \cite{xx} and references therein), our focus here is on 
The goal is twofold: (1) to estimate the unknown communication signal ${\bf S}_\text{soi}$ for data recovery, and (2) to estimate the reflected SI channel ${\bf H}_\text{si-c}$ for environmental sensing. 
%The communication performance is measured by the estimation accuracy of the SOI signal, ${\bf S}_\text{soi}$, and the sensing performance is characterized by the estimation accuracy of the SI channel ${\bf H}_\text{si-c}$. 
We treat the sum ${\bf H}_\text{soi}$ in \eqref{eq:H-SOI} as the overall channel coefficient of the SOI. 
Given that both the SI signal ${\bf S}_\text{si}$ and the SI direct channel ${\bf H}_\text{si-c}$ are known, we aim to estimate ${\bf H}_\text{si-r}$, the SI reflection channel, as well as the communication signal ${\bf S}_\text{soi}$.  %which is critical for characterizing the sensing performance. 

\subsection{Blind Source Separation for Sensing}
The FastICA algorithm exploits non-Gaussianity in the received signal mixture to estimate the independent source components iteratively. The algorithm follows these key steps:
\begin{enumerate}
    \item Preprocessing: Centering and whitening the received signal  ${\bf R}$  to eliminate correlations.
    \item Iterative Estimation: Maximize negentropy to extract statistically independent sources.
    \item Source Recovery: Estimate the unknown channel $\hat{\bf H}_\text{si-r}$ and signal $\hat{\bf S}_\text{soi}$ from the separated components.
    %Estimation of ${\bf H}_\text{si-r}$ and ${\bf S}_\text{soi}$: The estimated components corresponding to reflected SI and the information-carrying signal are extracted. Also,  $\hat{\bf H}_\text{si-r}$ and $\hat{\bf S}_\text{soi}$ are estimated.
\end{enumerate}
\subsection{Performance Metrics}

We evaluate the quality of channel estimation using the ergodic linear minimum mean squared error (ELMMSE), which quantifies the average estimation error of the sensed channel: 
\begin{align}
   \text{ELMMSE} = \mathbb{E}[ \|{\bf H}_\text{si-r} - \hat{ {\bf H}}_\text{si-r} \|^2 ].
\end{align}  
The communication performance is quantified via the 
 signal-to-residual-error ratio (SRER) of the extracted communication signal: 
\begin{align}
   \text{SRER} = \frac{ \mathbb{E}[ \|  {\bf S}_\text{soi} \|^2  ] }{\mathbb{E}[ \|  {\bf S}_\text{soi} - \hat{{\bf S}}_\text{soi} \|^2  ]  }.
\end{align} 
A higher SRER implies more effective signal separation and cleaner communication signal extraction, directly reflecting the system’s communication performance under the JCAS paradigm. 
Additionally, we can assess the ELMMSE of the overall SOI channel ${\bf H}_\text{soi}$ as
\begin{align}
\text{ELMMSE} = \mathbb{E}[ \|{\bf H}_\text{soi} - \hat{ {\bf H}}_\text{soi} \|^2 ],
\end{align}
which accounts for both the direct transmission channel between the transmitter and receiver, as well as the reflected and scattered components from the surrounding environment. This CSI can be leveraged to optimize the transmission schemes (e.g., precoder design in MIMO systems and optimizing the transmitted signal ${\bf S}_\text{soi}$), and to maximize the signal-to-interference-plus-noise ratio (SINR) 
\begin{align*}
   \text{SINR} = \frac{ \mathbb{E}[ \|  {\bf H}_\text{soi} {\bf S}_\text{soi} \|^2  ] }{\mathbb{E}[ \|  {\bf H}_\text{si} {\bf S}_\text{si}  \|^2  ]  + \mathbb{E}[ \|{\bf N} \|^2  ] }.
   \end{align*}

In Section \ref{sec:simulation}, we also track the number of iterations required for the FastICA algorithm to converge, which provides insight into the computational complexity and algorithmic efficiency of the proposed BSS approach under different signal block sizes.
By utilizing these metrics, we provide a comprehensive evaluation framework that jointly characterizes sensing accuracy, communication quality, and transmission efficiency in IBFD MIMO systems.   
%The signal-to-interference-plus-noise ratio (SINR) is 
%\begin{align}
 %   \frac{ |{\bf H}_\text{soi}^H P_\text{soi}|^2 }{|{\bf H}_\text{si}^H P_\text{si}|^2  + 1}
%\end{align}

%After obtaining the estimated 
%The BSS formulation 
%\begin{align}
%    {\bf X} = {\bf A S}
%\end{align}

\begin{figure*}[h]
\centering
\begin{subfigure}[b]{0.465\textwidth}
       \centering
        \includegraphics[width=1.05\linewidth]{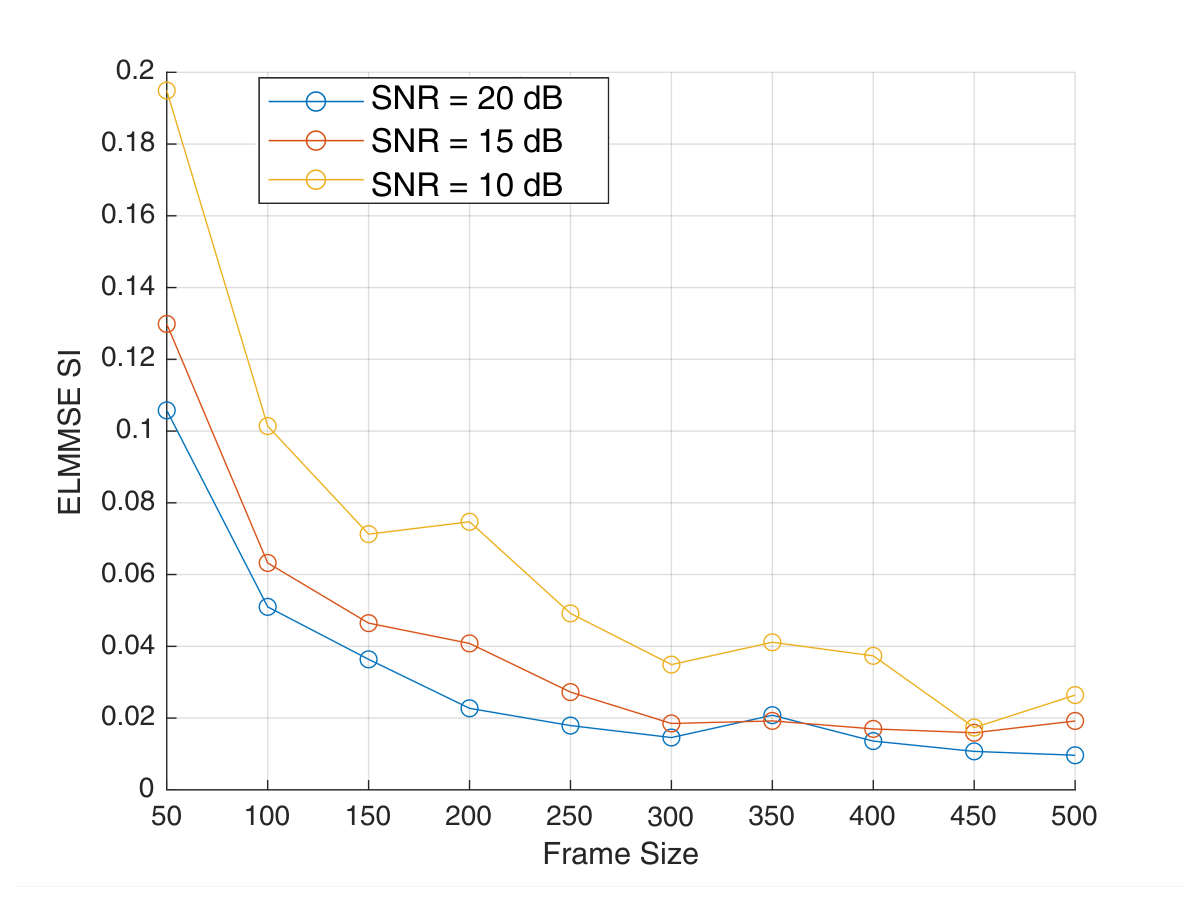} 
        \caption{ELMMSE of  channel sensing $\hat{\bf  H}_\text{si-r}$ v.s. frame size.}
        \label{fig:ELMMSE-SI}
    \end{subfigure}
    \hfill
\begin{subfigure}[b]{0.465\textwidth}
        \centering
        \includegraphics[width=1.05\linewidth]{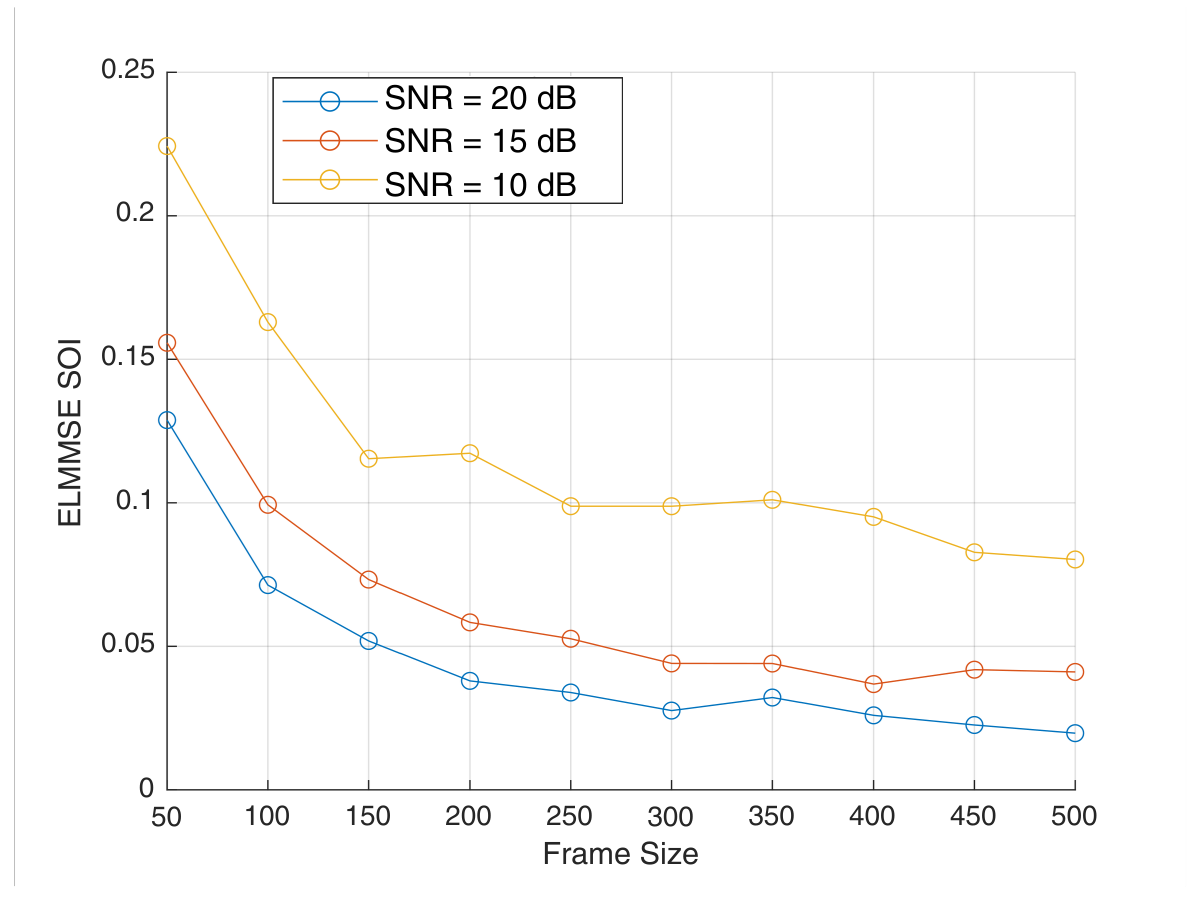} 
\caption{ELMMSE of  channel sensing $\hat{\bf  H}_\text{soi}$ v.s. frame size.}
     \label{fig:ELMMSE-SOI}
    \end{subfigure}
    \hfill
\begin{subfigure}[b]{0.465\textwidth}
        \centering
        \includegraphics[width=1.05\linewidth]{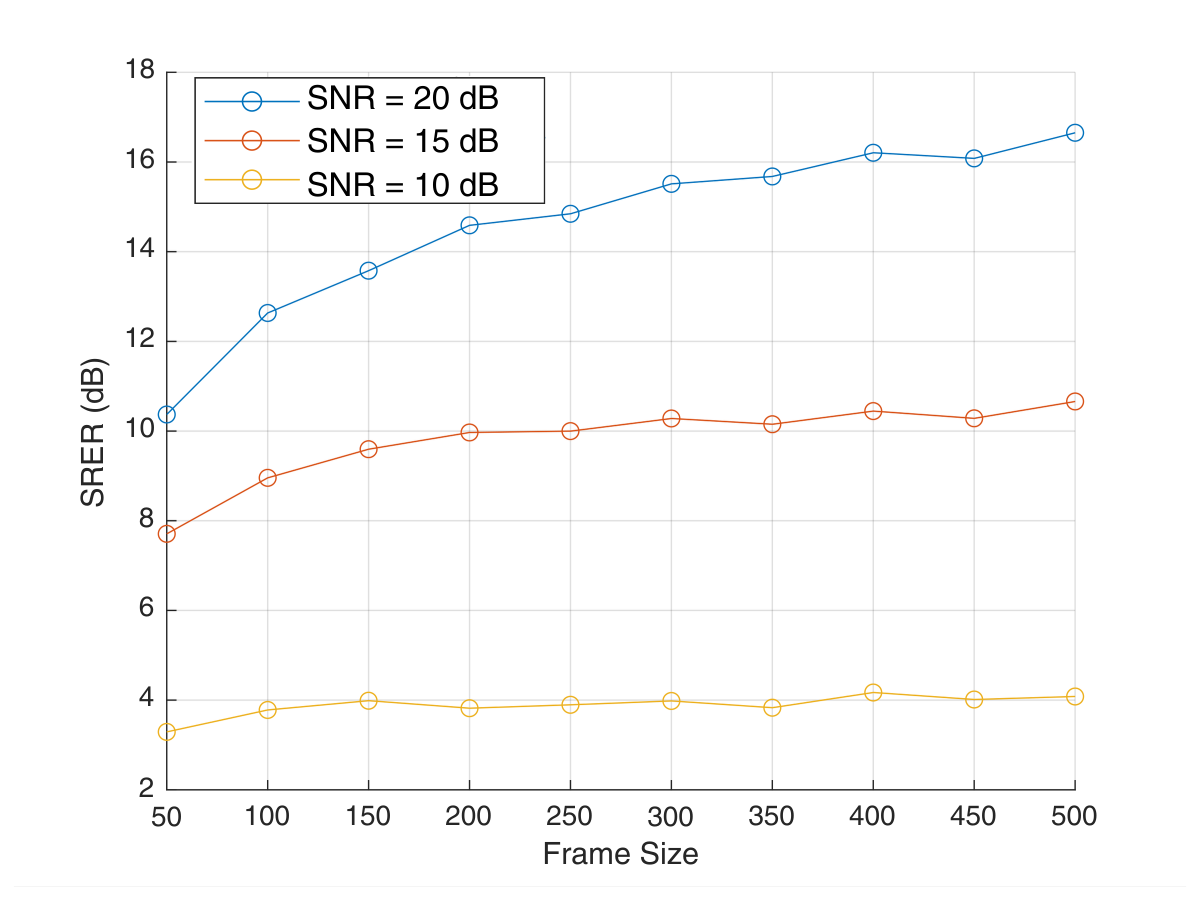} 
        \caption{SRER for SOI in dB v.s. frame size.}
        \label{fig:SRER}
    \end{subfigure}
    \hfill
\begin{subfigure}[b]{0.465\textwidth}
        \centering
        \includegraphics[width=1.05\linewidth]{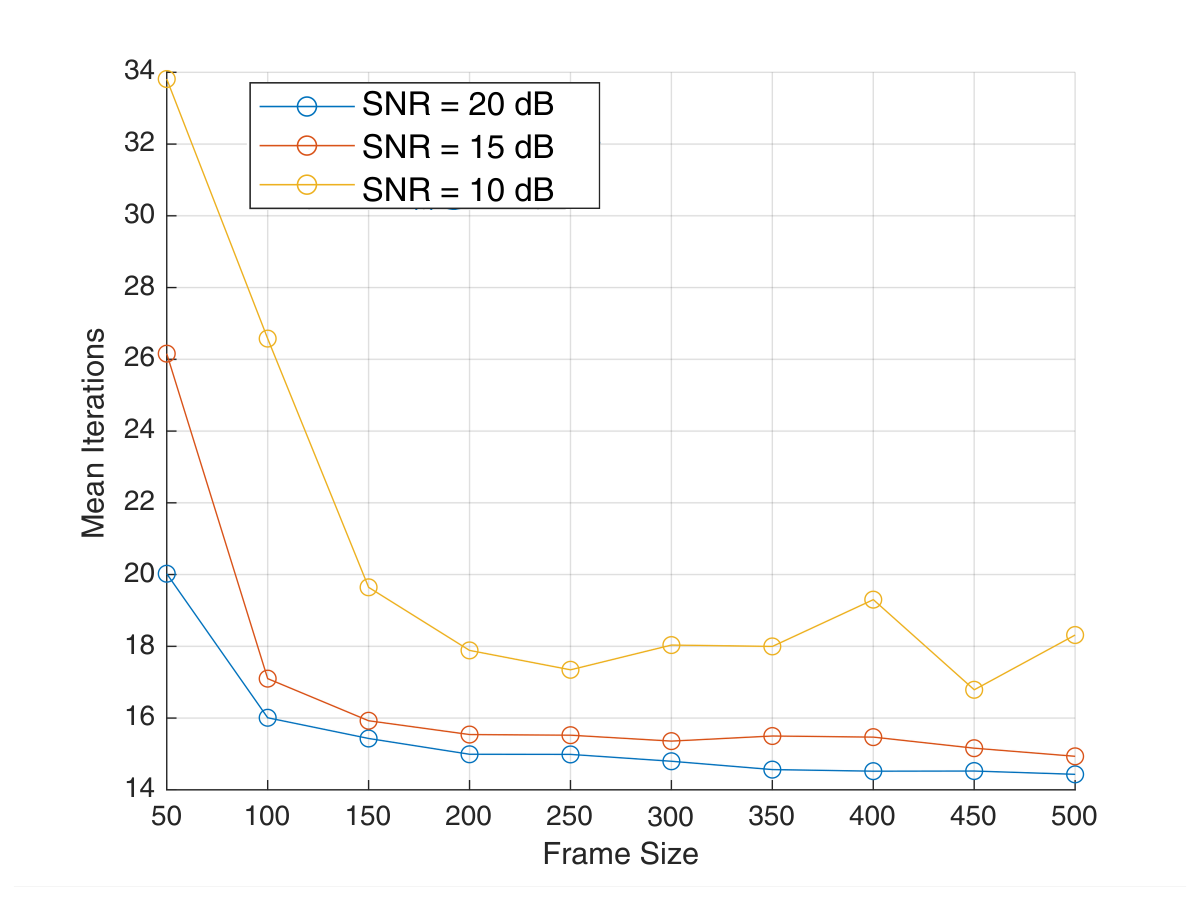}  
    \caption{Iterations required for convergence v.s. frame size.}
     \label{fig:Convergence-Iterations}
    \end{subfigure}
    \caption{BSS-enabled JCAS performance}
    \label{fig:performance-all}
\end{figure*}

\section{Simulation and Discussion}
\label{sec:simulation}

\subsection{Simulation Setup}
In this section, we evaluate the sensing and communication performance of the proposed system through numerical simulations based on a practical system model. Each node is equipped with two antennas. The reflected channel ${\bf H}_\text{si-r}$ is modeled as a Rician channel with ${\bf H}_\text{si-r} \sim \text{Rice}(x, y)$, capturing the presence of a dominant line-of-sight (LOS) component alongside multipath. In contrast, the overall communication channel ${\bf H}_\text{soi}$ is modeled as Rayleigh distributed, representing rich-scattering, non-LOS propagation environments. %In this section, we evaluate the sensing and communication performance via some numerical examples under a practical system model setup. In particular, we consider that each node is equipped with 2 antennas. The reflected channel ${\bf H}_\text{si-r}$  follows a Rician distribution where ${\bf H}_\text{si-r} \sim \text{Rice}(x, y)$. The overall communication channel ${\bf H}_\text{soi}$ follows a Rayleigh distribution, representing rich scattering without a dominant LOS path. 

The transmitted signals ${\bf S}_\text{si}$ and ${\bf S}_\text{soi}$ are BPSK modulated, with each symbol randomly chosen from $\{-1, +1\}$. The combined source matrix is defined as $\mathbf{S} = \begin{bmatrix} \mathbf{S}_{\text{si}}; \mathbf{S}_{\text{soi}} \end{bmatrix} \in \mathbb{R}^{4 \times N}$, where $N$ denotes the frame size (signal processing block length). The received signal is generated by linearly mixing the source signals through the corresponding channel matrices and adding Gaussian noise.

%${\bf S}_\text{si}$ and ${\bf S}_\text{soi}$ are generated as BPSK signals, where each symbol is randomly selected from $\{-1, +1\}$. The complete source matrix is $\mathbf{S} = \begin{bmatrix} \mathbf{S}_{\text{si}}; \mathbf{S}_{\text{soi}} \end{bmatrix} \in \mathbb{R}^{4 \times N}$ where $N$ denotes the signal processing block size (frame size). The signal at the receiver antenna is generated by mixing the sources with the channels and adding noise. 

Noise samples are generated from a zero-mean Gaussian distribution with variances $\sigma^2 = 0.01$, $0.0316$, and $0.1$, corresponding to signal-to-noise ratios (SNRs) of 20 dB, 15 dB, and 10 dB, respectively. These SNR values reflect low, moderate, and high noise environments. The FastICA algorithm is then applied to recover the sources, using kurtosis as the objective function to be maximized and a convergence threshold of $\epsilon = 10^{-6}$. Each simulation configuration is repeated over 1000 independent trials, and the average results are reported. The frame size is varied from 50 to 500 in increments of 50 to study performance across a broad range of processing lengths.

\subsection{Simulation Results}
Fig.~\ref{fig:performance-all} illustrates the system’s performance as a function of frame size and SNR level. The subfigures correspond to different SNR settings, 20 dB, 15 dB, and 10 dB, demonstrating the system’s robustness under varying noise conditions. %Fig. \ref{fig:performance-all} illustrates the proposed system’s performance under varying frame sizes and noise conditions. Each subplot in Fig. \ref{fig:performance-all} represents a different signal-to-noise ratio (SNR): 20 dB, 15 dB, and 10 dB, which correspond to low, medium, and high noise environments to test the robustness of this system in different conditions. 

Fig.\ref{fig:ELMMSE-SI} presents the evolution of the ELMMSE for the SI channel ${\bf H}_\text{si-r}$ as frame size increases. The results show a consistent decrease in estimation error, highlighting the benefit of larger frame sizes in improving sensing accuracy. A similar trend is observed in Fig.\ref{fig:ELMMSE-SOI} for the communication channel ${\bf H}_\text{soi}$. In both cases, lower SNRs lead to higher ELMMSE values, indicating increased difficulty in accurate channel estimation.
%Fig. \ref{fig:ELMMSE-SI} presents how the ELMMSE of the sensed channel for the SI signal ${\bf H}_\text{si-r}$ varies as the frame length changes, where the ELMMSE steadily decreases with increasing frame size, indicating that larger frame sizes lead to more accurate sensing of the SI signal channel. 
%Fig. \ref{fig:ELMMSE-SOI} shows a similar trend for the channel of communication signal of interest, ${\bf H}_\text{soi}$. The ELMMSE associated with the sensing of ${\bf H}_\text{soi}$ channel also declines with increasing frame length. Furthermore, varying SNR values result in different ELMMSE, since a higher SNR correlates to a lower result and a lower SNR correlates to a higher result for both Fig. \ref{fig:ELMMSE-SI} and Fig. \ref{fig:ELMMSE-SOI}.
Fig.~\ref{fig:SRER} shows the SRER in decibels, which reflects the quality of recovered communication signals. SRER improves steadily with frame size, particularly under higher SNRs. At 10 dB SNR, the SRER improvement is less pronounced due to strong noise.
%Fig.~\ref{fig:Convergence-Iterations} depicts the average number of iterations required by FastICA to converge. The number of iterations decreases with increasing frame size, especially at high SNR, where source separation becomes easier. At lower SNRs, convergence is slower due to the increased difficulty in distinguishing signal components amidst noise.
%The SRER expressed in decibels (dB), shown in Fig. \ref{fig:SRER}, quantifies the estimation accuracy of the recovered communication signal. The curve starts with relatively low SRER values at small frame sizes, and then exhibits a noticeable upward trend as the frame length increases. At a lower SNR of 10 dB, the SRER does not improve much since the noise is very dominant.
Fig. \ref{fig:Convergence-Iterations} depicts the number of iterations required by the FastICA algorithm to converge. For short frame sizes, even a modest increase in the frame size leads to a significant improvement in convergence speed. As the frame size continues to increase, the number of iterations required for convergence gradually decreases at a slower rate, forming a gently sloping trend. The average number of iterations required for convergence is directly influenced by the SNR, i.e., higher SNR levels lead to faster convergence with fewer iterations, while lower SNR conditions increase the number of iterations due to the greater difficulty in signal separation under noisy environments.

The performance improvements observed across all evaluated metrics are attributed to the increased accuracy in signals' statistical property estimation for the FastICA algorithm when a greater number of data samples are available. As the frame size grows, more accurate signal separation and channel estimation are achieved, together with faster convergence speed.
However, beyond a certain threshold, approximately 350 symbols per frame, the rate of improvement begins to plateau, indicating diminishing returns with further increases in frame size. This insight is informative for the choice of appropriate frame sizes, especially for communication systems operating in rapidly changing channel environments, when smaller frame sizes may be advantageous due to the performance degradation caused by changes in channel parameters within each frame.  

\subsection{Discussion on the BSS in IBFD MIMO JCAS}  
The complexity of the FastICA algorithm  is generally given by $\cO(m^3+ k m^2 N )$, where $m$ is the number of sources to be separated (in our model, this corresponds to the combined number of transmit antennas from both nodes), $N$ is the frame size, and $k$ is the number of iterations until convergence. In our simulated system, the number of antennas ($m=4$) is small, making the $\cO(m^3)$ term, associated with the initial whitening process, negligible. The dominant factor is the iterative estimation process. Our simulation result in Fig. \ref{fig:Convergence-Iterations} shows that the number of iterations for convergence is modest and decreases as the frame size increases from 50 to 200 symbols. For a frame size of 350 symbols, the algorithm converges in fewer than 18 iterations on average, even at a low SNR of 10 dB.  In practical terms, FastICA is well-suited for real-time implementation on modern digital signal processors (DSPs) or GPU platforms, particularly when antenna dimensions are modest (e.g., $m \leq 8$).

The integration of BSS techniques into IBFD MIMO systems for JCAS offers compelling advantages that align with the goals of next-generation wireless systems. One of the most significant benefits of such JCAS method is the elimination of dedicated sensing signals. Since the SI signal is already transmitted and fully known at the node, it can be repurposed for environmental sensing, transforming a traditionally detrimental component into a valuable asset. This not only simplifies system design but also reduces the overhead associated with conventional radar signal transmission. Another key advantage lies in the improved spectral efficiency that naturally arises from the JCAS paradigm. By enabling simultaneous communication and sensing within the same frequency band, the proposed IBFD systems avoid the need for additional bandwidth allocation, making more efficient use of the limited spectrum. Furthermore, since BSS methods operate by exploiting statistical independence of source signals, they are able to effectively separate sources without requiring explicit channel state information. In this work, we assume static channel conditions, leaving the dynamic scenarios for future study.

\section{Conclusion and Future Work}
\label{sec:conclusion}
In this work, we proposed a blind source separation (BSS)-based framework to enable joint communication and sensing in in-band full-duplex MIMO systems. Leveraging the fully known self-interference signal, our approach repurposes the self-interference signal for environmental sensing, eliminating the need for additional sensing waveforms. We formulated the signal model as a BSS problem and applied the FastICA algorithm to separate the self-interference signal from the signal of interest, with simultaneous channel estimation. 
Simulation results confirm that the proposed method is fully capable of joint sensing and communication. Also, it was found that while the system performance improves with increasing frame size, the improvement plateaus beyond a certain frame size.

In this study, we focused on a simplified scenario with BPSK source signals. Our framework can be naturally extended to more practical modulation schemes such as QPSK and QAM, where the source signals are complex-valued. This extension will necessitate the use of complex-valued BSS algorithms. Future work will also explore the application of the proposed method to dynamic time-varying channels, which pose many challenges to wireless communication systems. In such scenarios, while longer frames enable more accurate estimation of signals' statistical properties (which translates to improved signal separation performance in static channel conditions), long frame sizes cause degradation of BSS performance in fast-changing channel conditions, because the mixing matrix changes within each individual frame. This issue not only affects signal separation accuracy, but may also cause some BSS algorithms (such as Fast-ICA) not able to converge. Therefore, the selection of the frame size employed in BSS algorithms becomes critical, and BSS algorithms specifically designed for operation in dynamic mixing processes are desirable. For such scenarios, we will explore the integration of reinforcement learning-based mechanisms to adaptively select optimal frame sizes based on real-time channel dynamics.

\bibliographystyle{IEEEtran}
\bibliography{ref}
\end{document}